 \documentclass[final,3p,times,twocolumn]{elsarticle}

\usepackage{amssymb}
\usepackage{amsmath}
\usepackage{hyperref}% Creates hyperlinks

\usepackage{graphicx}% Include figure files
\usepackage{dcolumn}% Align table columns on decimal point
\usepackage{bm}% bold math
\usepackage{epstopdf}% pdflatex command active even for eps figures
\usepackage{bm,color,breakurl}

\journal{Journal of Magnetism and Magnetic Materials}

	\usepackage{fancyheadings}
\renewcommand{\thispagestyle}[1]{} % do nothing

\begin{document}



\title{Magnetocaloric and electrocaloric properties of the Hubbard pair cluster}


\author[a1]{K. Sza\l{}owski\corref{cor1}}
\ead{karol.szalowski@uni.lodz.pl}
\ead[url]{https://orcid.org/0000-0002-3204-1849}
%\thanks{}
\author[a1]{T. Balcerzak}
\ead[url]{https://orcid.org/0000-0001-7267-992X}
\ead{tadeusz.balcerzak@uni.lodz.pl}
%\thanks{}
%\thanks{}
\address[a1]{Department of Solid State Physics, Faculty of Physics and Applied Informatics,\\
University of \L\'{o}d\'{z}, ulica Pomorska 149/153, 90-236 \L\'{o}d\'{z}, Poland}

\cortext[cor1]{Corresponding author}

\date{\today}

\begin{abstract}
The paper contains the discussion of the magnetocaloric and electrocaloric effect in a model  dimer (pair cluster). The system of interest is modelled with a Hubbard Hamiltonian including the external electric and magnetic field. The thermodynamics of such pair is described exactly, on the grounds of the grand canonical ensemble, focusing on the half-filling of energy states.  The quantities of interest, such as magnetic entropy, magnetic specific heat as well as isothermal entropy change resulting from the variation of either electric or magnetic field and appropriate Gr\"uneisen ratios are calculated and discussed in a wide range of external fields. The importance of singlet to triplet transition for the observed behaviour is emphasized. The ranges of direct and inverse caloric effects are found and the manifestations of the magnetoelectric phenomena are described. In particular, the tunability of the magnetocaloric effect with electric field as well as tunability of the electrocaloric effect with magnetic field are demonstrated.
\end{abstract}


\maketitle

%section 1
\section{Introduction}

The development of nanodevices stimulates strongly the search for novel approaches to  refrigeration in relevant scale, ranging from nano- to mesoscale, exploiting a plethora of physical phenomena \cite{Giazotto2006,Muhonen2012,Ziabari2016}. Among various strategies adopted to achieve the goal of on-chip cooling, one of the successful approaches is based on magnetocaloric effect (MCE)\cite{Pecharsky2001a,Franco2018}, manifesting itself for example in a form of temperature drop during adiabatic demagnetization \cite{Belo2019}. This principle has been demonstrated and used for effective on-chip cooling on the basis of such subsystems as single magnetic ions \cite{Ciccarelli2016}, thin magnetic films \cite{Bradley2017} or nuclear magnetic moments \cite{Palma2017,Yurttagul2019,Sarsby2020}. Analogous mechanism has already been observed in molecular nanomagnets \cite{Sharples2014}. Another, less frequently investigated caloric effect connected with the variability of the external electric fieldis electrocaloric effect (ECE) \cite{Kutnjak1999,Ozbolt2014}. Both effects give hopes for efficient solid state-based cooling \cite{Manosa2013} and motivate constant quest for novel materials and concepts, including especially quantum materials \cite{Reis2020}.

The finite cluster nanosystems exhibit typically quantum level crossings \cite{Waldmann2007,Furrer2013a} - the points in which the ground state of the system changes when some control parameter (like the external electric or magnetic field) is varied. Their presence manifests itself in the experiment, for example, as a rapid change of the total magnetization of the system at a certain critical value of the field. However, the accidental state degeneracy at the quantum level crossing point causes also a residual entropy to emerge exactly for the critical field. This entropy has necessarily different value than the entropy at each side of the quantum level crossing point (which may be either zero when the ground state is non-degenerate or positive if it is degenerate). As a consequence, the ground-state (residual) entropy exhibits a discontinuous behaviour as a function of the field at the quantum level crossing point. The temperature would tend to smear that dependence, nevertheless, such a behaviour of the entropy makes it particularly sensitive to the external field in the vicinity of the quantum level crossing. This behaviour can be utilized to maximize the caloric effect corresponding to the external field causing the quantum level crossing. 

The simplest magnetic systems with spin-spin coupling are spin dimers. In case of antiferromagnetic coupling, they exhibit a singlet-to-triplet transition (quantum level crossing) when the magnetic field is increased. The dimer structure can arise naturally in molecular magnets, making them a highly interesting class of materials. The mentioned phenomenon has been studied in the context of the magnetocaloric effect, for example in (coupled) Cu-based dimers with spin 1/2 \cite{Chakraborty2015,Brambleby2017,Chakraborty2019} or Ni-based dimers with spin 1 \cite{Tarasenko2020}. It can be mentioned that also rotational magnetocaloric effect utilizing magnetic anisotropy has been studied in dimer systems based on Dy and Gd ions \cite{Lorusso2016}. The phenomenon has been also found and discussed theoretically in various magnetic cluster systems, to mention such examples as the calculations for anisotropic Heisenberg polyhedra \cite{Karlova2017e}, Ising tetrahedra \cite{Mohylna2019d}, edge-sharing tetrahedra and octahedra \cite{Zad2019a} or triangular lattice-based Ising nanoclusters \cite{Zukovic2015d,Mohylna2020} and other clusters \cite{Haldar2020}. 

In order to enrich the number of degrees of freedom in the studied system and include two external fields - the magnetic and electric one - a natural choice is focusing the interest on a Hubbard dimer (pair). Such nanosystem exhibits an interplay between charge and spin response to the external fields, being a natural candidate system to exhibit pronounced magnetoelectric phenomena. Some properties of the system, like the chemical potential, magnetic and electric polarization and susceptibilities were studied by us in Refs.~\cite{Balcerzak2017c,Balcerzak2018d,Balcerzak2018e}
It should be mentioned that the thermodynamics of such system can be described exactly. The Hubbard dimer has been studied also in the context of symmetries \cite{Cerrato2019}, density functional theory \cite{Carrascal2015a,Ullrich2018b}, spectral function \cite{Vanzini2018}, integrals of motion \cite{Wortis2017a}, two-orbital model \cite{Amendola2015}, orbital degeneracy \cite{Spalek1979a} or the extended version of the Hubbard model \cite{Chen1978,Iglesias1997b} including the electron-phonon couplings within Hubbard-Holstein model on a dimer \cite{Acquarone1998}. Other cluster-based Hubbard nanostructures have also been studied \cite{Callaway1990,Pastor1994,Lopez-Urias2005}, to mention especially those like cube \cite{Callaway1987,Schumann2010a,Szalowski2018c}, triangle \cite{Schumann2007a,Juliano2019} and tetrahedron \cite{Schumann2007a} or a finite chain \cite{Noce1997a,Hancock2005a,Hancock2014a}. 

In the paper we characterize exactly the magnetocaloric and electrocaloric effect in Hubbard dimer (pair), exploiting the plethora of phenomena caused by the simultaneous presence of electric and magnetic field. In the next section \ref{theory} we sketch the theoretical formalism used to characterize the thermodynamics of the Hubbard dimer, based on the grand canonical ensemble. In the following part \ref{num} we present and discuss the results of the extensive numerical calculations focused on the magneto- and electrocaloric effect. The  final remarks are drawn in the section \ref{final}.

%section 2
\section{\label{theory}Theoretical model}

The Hubbard Hamiltonian for the pair of atoms $(a,b)$ embedded in the external magnetic and electric fields is of the form:
\begin {align}
\mathcal{H}_{a,b}=&-t\sum_{\sigma=\uparrow,\downarrow}\left( c_{a,\sigma}^+c_{b,\sigma}+c_{b,\sigma}^+c_{a,\sigma} \right)\nonumber\\&+U\left(n_{a,\uparrow}n_{a,\downarrow}+n_{b,\uparrow}n_{b,\downarrow}\right)\nonumber\\&
-H\left(S_a^z+S_b^z\right) -V\left(n_{a}-n_{b}\right),
\label{eq1}
\end {align}
where $t>0$ is the hopping integral, and $U\ge0$ is on-site Coulomb repulsion energy. The external uniform magnetic field with magnitude $H^z$ is introduced by the parameter $H$, namely $H=-g\mu_{\rm B}H^z$. The parameter $V$ stands for the electrostatic potential of the uniform electric field oriented along the line joining both atoms, in such a way that $V=V_{a}=-V_{b}$. It is related to the electric field $E$ by the formula $V=E|e|d/2$, where $d$ is the interatomic distance and $e$ is the electron charge.

The creation ($c_{\gamma,\sigma}^+$) and annihilation ($c_{\gamma,\sigma}$) operators for site $\gamma = a,b$ and spin state $\sigma=\uparrow,\downarrow$ can be used to define the corresponding occupation number operators $n_{\gamma,\sigma}$, namely:
\begin {equation}
n_{\gamma,\sigma}=c_{\gamma,\sigma}^{+}c_{\gamma,\sigma}.
\label{eq2}
\end {equation}
With the help of $n_{\gamma,\sigma}$ the total occupation number operators at the site $\gamma = a,b$ are given by:
\begin {equation}
n_{\gamma}=n_{\gamma,\uparrow}+n_{\gamma,\downarrow}.
\label{eq3}
\end {equation}
Moreover, the spin operators $S_{\gamma}^z$ in Eq.(\ref{eq1}) are defined as follows:
\begin {equation}
S_{\gamma}^z=\left(n_{\gamma,\uparrow}-n_{\gamma,\downarrow}\right)/2.
\label{eq4}
\end {equation}
We consider the Hubbard pair as an open system, being able to exchange the electrons with its neighbourhood.  For instance, this could be a situation where the pair cluster is placed on the metallic substrate serving as the electronic reservoir or a situation where the pair is coupled to the electrodes (gates). Taking into account the possible fluctuations of the number of electrons, the equilibrium thermodynamics of such open system is properly described by the grand canonical ensemble. In this formalism the Hamiltonian (\ref{eq1}) is extended by adding the term  $-\mu \left(n_a+n_b\right)$, where $\mu$ is the chemical potential.

The exact analytical diagonalization of the extended Hamiltonian has been performed in Ref.~\cite{Balcerzak2017c}. As a result, a set of 16 eigenenergies and corresponding eigenstates has been found, which enabled determination of the grand partition function $\mathcal{Z}_{a,b}$:
\begin {align}
\mathcal{Z}_{a,b}&= {\rm Tr}_{a,b} \,\exp \lbrace -\beta \left[\mathcal{H}_{a,b}-\mu\left(n_a+n_b\right)\right]\rbrace
\nonumber\\&=\sum_{i=1}^{16} \exp \left(-\frac{t}{k_{\rm B}T}E_i \right),
\label{eq5}
\end {align}
where $E_i$ are the normalized energy eigenvalues given in the Appendix B of Ref.~\cite{Balcerzak2017c}.
The grand potential, $\Omega_{a,b}$, of the open system is then given by:
\begin {equation}
\Omega_{a,b}=-k_{\rm B}T \ln \mathcal{Z}_{a,b},
\label{eq6}
\end {equation}
and it enables the calculation of all thermodynamic properties in equilibrium.

On the other hand, the statistical properties can be found from the statistical operator $ \rho_{a,b}$:
\begin {equation}
\rho_{a,b}=\frac{1}{\mathcal{Z}_{a,b}}\, \exp \left\{ -\beta \left[\mathcal{H}_{a,b}-\mu\left(n_a+n_b\right)\right]\right\},
\label{eq7}
\end {equation}
which can be constructed in a diagonal form on the basis of the diagonalized pair Hamiltonian $\mathcal{H}_{a,b}$. With the help of $ \rho_{a,b}$ the statistical averages of arbitrary quantum mechanical operators can be calculated. In particular, averaging of operators $n_{\gamma}$, and $S_{\gamma}^z$, which are defined by Eqs.~(\ref{eq3}) and (\ref{eq4}), respectively, can be performed. Namely:
\begin {equation}
\left<n_{\gamma} \right>={\rm Tr}_{a,b} \left[ \left(n_{\gamma,\uparrow}+n_{\gamma,\downarrow}\right)\; \rho_{a,b} \right],
\label{eq8}
\end {equation}
and
\begin {equation}
\left<S_{\gamma}^z \right>={\rm Tr}_{a,b} \left[ \frac{1}{2}\left(n_{\gamma,\uparrow}-n_{\gamma,\downarrow}\right)\; \rho_{a,b} \right].
\label{eq9}
\end {equation}
For completeness of the method, the chemical potential $\mu$ can be self-consistently determined from the relationship:
\begin {equation}
\left<n_a\right>+\left<n_b\right>=-\left(\frac{\partial \Omega_{a,b}}{\partial \mu}\right)_{T,H,V}.
\label{eq10}
\end {equation}
For studies of the magnetocaloric and electrocaloric effects the entropy $S$ of the system is a crucial quantity. The entropy as a function of $T$, $H$ and $E$ is defined by:
\begin {equation}
S\left(T,H,E\right)=-\left(\frac{\partial \Omega_{a,b}}{\partial T}\right)_{H,E},
\label{eq11}
\end {equation}
where the external field parameters $H$ and $E$ are constant. 

Alternatively, the entropy can be expressed as:
\begin {equation}
S\left(T,H,E\right)=\frac{\left\langle \mathcal{H}_{a,b}-\mu\left(n_a+n_b\right)\right\rangle-\Omega_{a,b}}{T}.
\label{eq11b}
\end {equation}
The caloric effects, which can manifest themselves by the heat flow between the system and its environment under the external field change can be quantified with the help of the isothermal entropy changes $\Delta S_T$. For the magnetocaloric effect we define 
\begin{equation}
\Delta S_T^{MCE}=S(T,H=0,E)-S(T,H,E),
\label{deltaSH}
\end{equation}
i.e., $\Delta S_{T}$ is the isothermal change of the entropy corresponding to the jump of magnetic field from $H=0$ to $H>0$, whereas the electric field parameter $E$ is constant. Analogously, the electrocaloric effect is described by
\begin{equation}
\Delta S_T^{ECE}=S(T,H,E=0)-S(T,H,E),
\label{deltaSE}
\end{equation}
where the isothermal entropy change corresponds to the jump of the electric field from $E=0$ to $E>0$, whereas the magnetic field is constant.

The known entropy of the system can also be exploited for calculation of the heat capacity, $C_{H,E}$. For the constant external field parameters $H$ and $E$, the heat capacity of the Hubbard pair cluster (dimer) is then given by:
\begin {equation}
C_{H,E}=T\left(\frac{\partial S\left(T,H,E\right)}{\partial T}\right)_{H,E}=-T\left(\frac{\partial^2 \Omega_{a,b}}{\partial T^2}\right)_{H,E}.
\label{eq12}
\end {equation}
Applying the fluctuation-dissipation theorem, an alternative formula, particularly convenient for numerical calculations, can be derived in the following form:
\begin {equation}
C_{H,E}=\frac{\left\langle \left[\mathcal{H}_{a,b}-\mu\left(n_a+n_b\right)\right]^2\right\rangle-\left\langle \mathcal{H}_{a,b}-\mu\left(n_a+n_b\right)\right\rangle^2}{k_{\rm B}T^2}.
\label{eq12b}
\end {equation}

Potentially interesting parameters quantifying the response of the system to the external field are Gr\"uneisen ratios. For the system embedded in external magnetic field $H$ and electric field $E$ two such parameters can be defined. Namely, a magnetic Gr\"uneisen ratio can be defined as \cite{Zhu2003, Szalowski2011a}:
\begin {equation}
\Gamma_H=-\frac{1}{C_{H,E}}\left(\frac{\partial M}{\partial T}\right)_{H,E},
\label{eq13}
\end {equation}
where $M=\left<S_{a}^z \right>+\left<S_{b}^z \right>$ is the total magnetization of the cluster. This quantity can be further expressed in the following forms:
\begin {equation}
\Gamma_{H}=\frac{1}{T}\left(\frac{\partial T}{\partial H}\right)_{S,E}=-\frac{1}{C_{H,E}}\left(\frac{\partial S}{\partial H}\right)_{T,E}.
\label{eq14}
\end {equation}

Moreover, in analogous manner, an electric Gr\"uneisen ratio can be defined as follows:
\begin {equation}
\Gamma_E=-\frac{1}{C_{H,E}}\left(\frac{\partial P}{\partial T}\right)_{H,E},
\label{eq15}
\end {equation}
where $P$ is the total electric polarization of the pair. Alternatively, it can be expressed as:
\begin {equation}
\Gamma_{E}=\frac{1}{T}\left(\frac{\partial T}{\partial E}\right)_{S,H}=-\frac{1}{C_{H,E}}\left(\frac{\partial S}{\partial E}\right)_{T,H}.
\label{eq16}
\end {equation}
The formulas given by  Eq.~\ref{eq14} and \ref{eq16} show a direct relation of Gr\"uneisen ratios both to the differential temperature change under adiabatic conditions and to the differential entropy change under isothermal conditions, thus proving the importance of these quantities for description of the caloric effects. Interestingly, the Gr\"{u}neisen ratio is expected to diverge at quantum phase transition points \cite{Zhu2003,Jafari2012} and presents an experimentally measurable quantity.

The numerical calculations based on the above formalism and aimed at description of the magneto- and electrocaloric effects for the Hubbard dimer will be presented in the next Section~\ref{num}.\\

%section 3
\section{\label{num}Numerical results and discussion}

The numerical results have been obtained on the basis of formalism outlined in previous section and are based on the exact diagonalization of the model. For most of the figures, the mean number of electrons in the cluster has been assumed as
$ x=\left(\left<n_a\right>+\left<n_b\right>\right)/2=1$, which corresponds to the half-filling condition for the energy states of the system. For such electron concentration, the chemical potential is independent on the external fields and temperature and equal to $\mu=U/2$. However, two figures were prepared to demonstrate the influence of the electron concentration $x$ on the thermodynamic parameters.

\begin{figure}[t]
\begin{center}
\includegraphics[width=1.1\columnwidth]{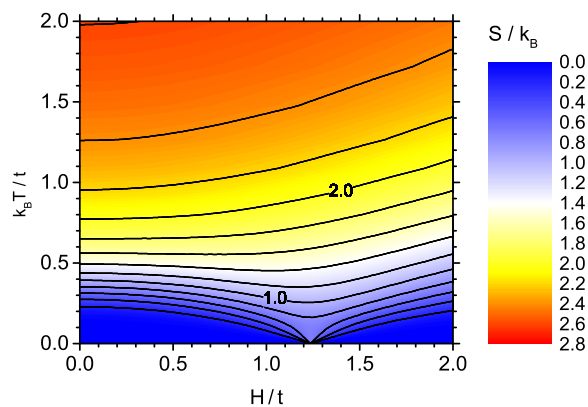}
\caption{\label{fig1}Density plot of normalized entropy as a function of the normalized temperature and magnetic field, for electric field $E|e|d/t=0.0$ and $U/t=2.0$. Isentropes are marked with solid lines.}
\end{center}
\end{figure}

We commence the discussion of the results from the most fundamental quantity for the caloric properties - the entropy of the system. In Fig.\ref{fig1}, the normalized entropy $S/k_{\rm B}$ is presented in the normalized magnetic field $H/t$ - normalized temperature $k_{\rm B}T/t$ coordinates as a density plot with contours. The electric field is absent in this case. The isentropes with increasing values correspond typically to increasing temperatures. For zero temperature, a characteristic point is seen, in which the isolines are concentrated. This point corresponds to the magnetic critical field $H_{c}$ in which the quantum level crossing takes place, as the system switches from a singlet state (occurring in lower magnetic fields) to a triplet state (occurring in higher magnetic fields) at zero temperature. The value of the critical field $H_{c}$, seen in Fig.\ref{fig1} for the normalized electric field $E|e|d/t=0$ and Hubbard on-site energy $U/t=2$, is in agreement with the phase diagram constructed by us in Ref.~\cite{Balcerzak2018e} (see Fig.~1 in Ref.~\cite{Balcerzak2018e}).

\begin{figure}[t]
\begin{center}
\includegraphics[width=1.1\columnwidth]{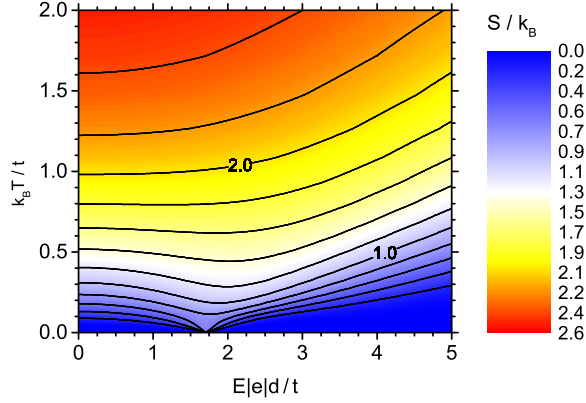}
\caption{\label{fig2}Density plot of normalized entropy as a function of the normalized temperature and electric field, for magnetic field $H/t=1.5$ and $U/t=2.0$. Isentropes are marked with solid lines.}
\end{center}
\end{figure}

\begin{figure}[t]
\begin{center}
\includegraphics[width=0.99\columnwidth]{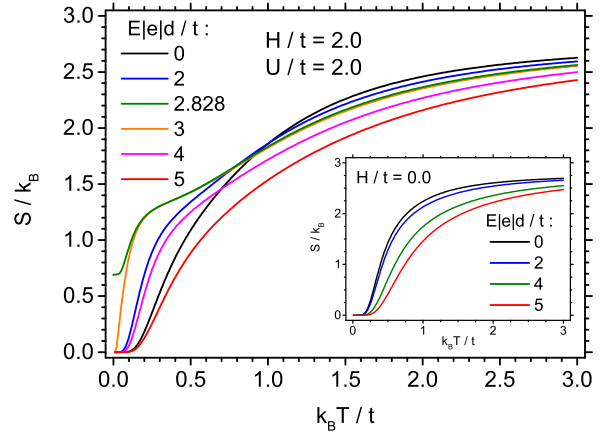}
\caption{\label{fig3}Dependence of the normalized entropy on the normalized temperature, for $U/t=2.0$ for various electric fields, for normalized magnetic field $H/t=2.0$ (main panel) and for $H/t=0.0$ (inset).}
\end{center}
\end{figure} 

\begin{figure}[t]
\begin{center}
\includegraphics[width=1.1\columnwidth]{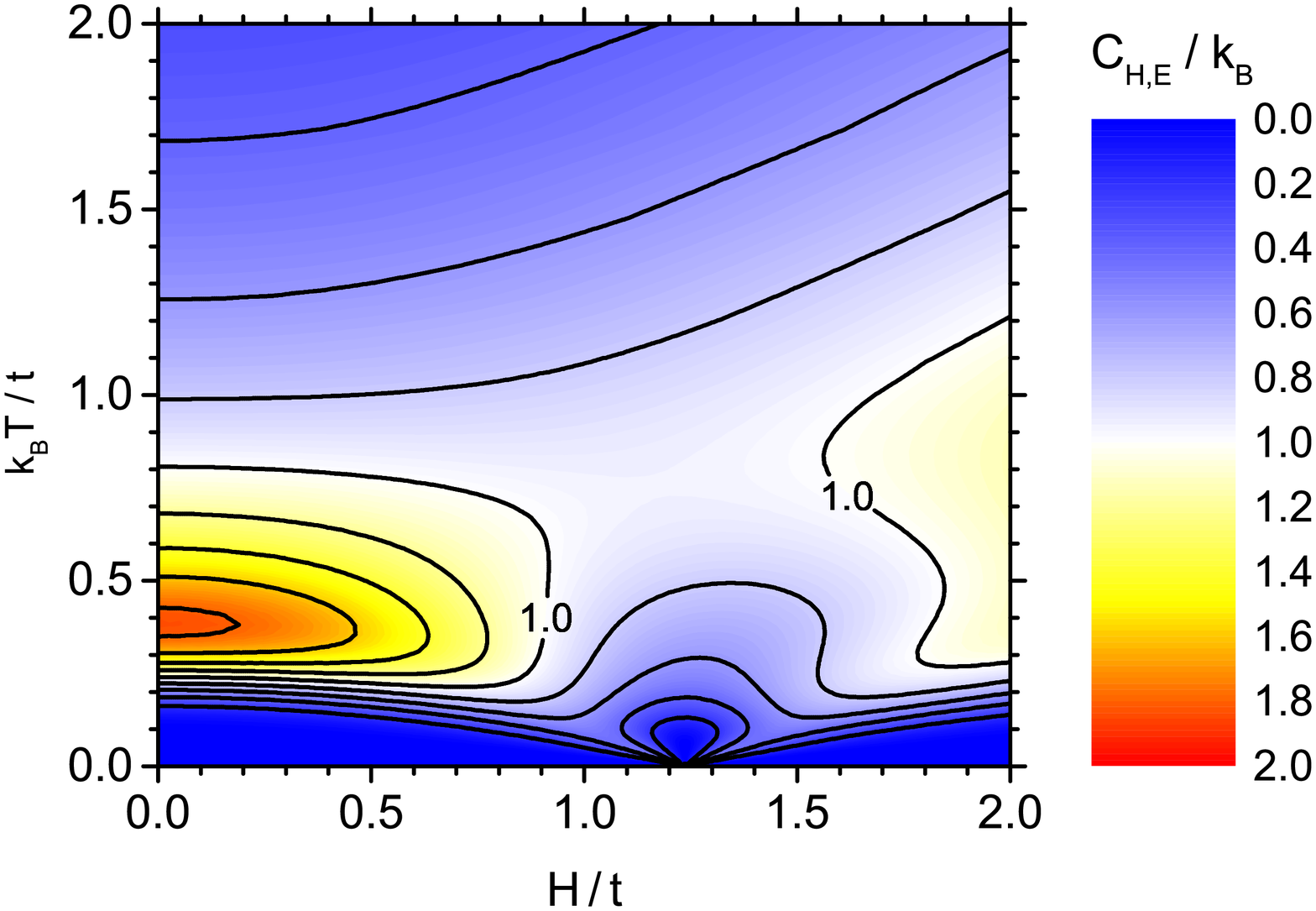}
\caption{\label{fig4}Density plot of normalized specific heat as a function of the normalized temperature and magnetic field, for electric field $E|e|d/t=0.0$ and $U/t=2.0$. Lines of constant specific heat are marked with solid lines.}
\end{center}
\end{figure}

To complement the picture presented in Fig.~\ref{fig1}, in Fig.\ref{fig2} the normalized entropy $S/k_{\rm B}$ is presented in the normalized electric field $E|e|d/t$ -  normalized temperature $k_{\rm B}T/t$ coordinates as a density plot with contours. The magnetic field is set to $H/t=1.5$. As before, the isentropes with increasing values correspond normally to increasing temperatures. Again, for zero temperature, a characteristic point is seen, in which the isolines are concentrated. This point corresponds to the electric critical field $E_{c}$ in which the quantum level crossing takes place and the system switches from triplet state (occurring in lower electric fields) to singlet state (occurring in higher electric fields). It can be verified that the value of the critical field $E_{c}$ presented in Fig.\ref{fig2} for the normalized magnetic field $H/t=1.5$ and Hubbard on-site energy $U/t=2$ is in agreement with the phase diagram found by us in Ref.~\cite{Balcerzak2018e}.  

It can already be noted that in the vicinity of the quantum level crossing the entropy becomes particularly sensitive to the external field - either magnetic or electric one. This fact is connected with the possibility of generating a significant entropy change with a limited change in the field, thus maximizing the caloric effects. Such a conclusion has been drawn, for example, in Ref.~\cite{Chakraborty2019} for the magnetocaloric effect in Heisenberg dimer, undergoing the singlet-triplet transition.

The cross-sections of the density plots permit the detailed tracking of the entropy variability as a function of a single control parameter. An example of such plot is Fig.~\ref{fig3}, where the entropy, $S/k_{\rm B}$, is plotted vs. dimensionless temperature  $k_{\rm B}T/t$ for several electric fields $E|e|d/t$ and for $U/t=2$. In the main panel the magnetic field is fixed at $H/t=2$, whereas in the inset the magnetic field is absent. In general, the entropy is an increasing function of temperature, and when $T \to \infty$ the entropy reaches the limit $S/k_{\rm B}= \ln 16 \approx 2.7726$. It means that all the 16 states of the Hubbard pair cluster, which have been specified in Ref.~\cite{Balcerzak2017c}, are occupied with equal probability. The electric field causes that approaching this limit is slightly harder. On the other hand, for $T \to 0$, when the system is in a pure ground state, either singlet or triplet one, the entropy goes to zero. However, for the electric critical field (the green curve labelled  by $E_{c}|e|d/t=2.828$), i.e., when the quantum level crossing takes place, the residual entropy remains. It results from degeneracy of two states (singlet and triplet) exactly at the phase transition point, and its value is $S/k_{\rm B}= \ln 2 \approx 0.6931$. In general, the residual entropy, $S/k_{\rm B}$ amounts to $\ln n$, where $n$ is a number of degenerate states occurring with the same ground state energy. In our case $n=2$, which corresponds to equilibrium coexistence of the singlet and triplet state exactly at the critical electric field. For $H/t=0$, in the inset, the system is in pure singlet state for all considered values of the electric field and the residual entropy does not emerge. It should also be noted that for $T \to 0$ the entropy curves exhibit a vanishing slope, thus not depending on the temperature, which reflects the 3rd law of thermodynamics.

\begin{figure}[t]
\begin{center}
\includegraphics[width=1.1\columnwidth]{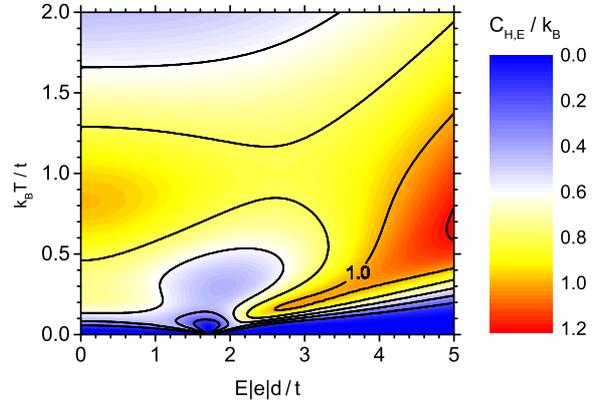}
\caption{\label{fig5}Density plot of normalized specific heat as a function of the normalized temperature and electric field, for magnetic field $H/t=1.5$ and $U/t=2.0$. Lines of constant specific heat are marked with solid lines.}
\end{center}
\end{figure}

\begin{figure}[t]
\begin{center}
\includegraphics[width=0.99\columnwidth]{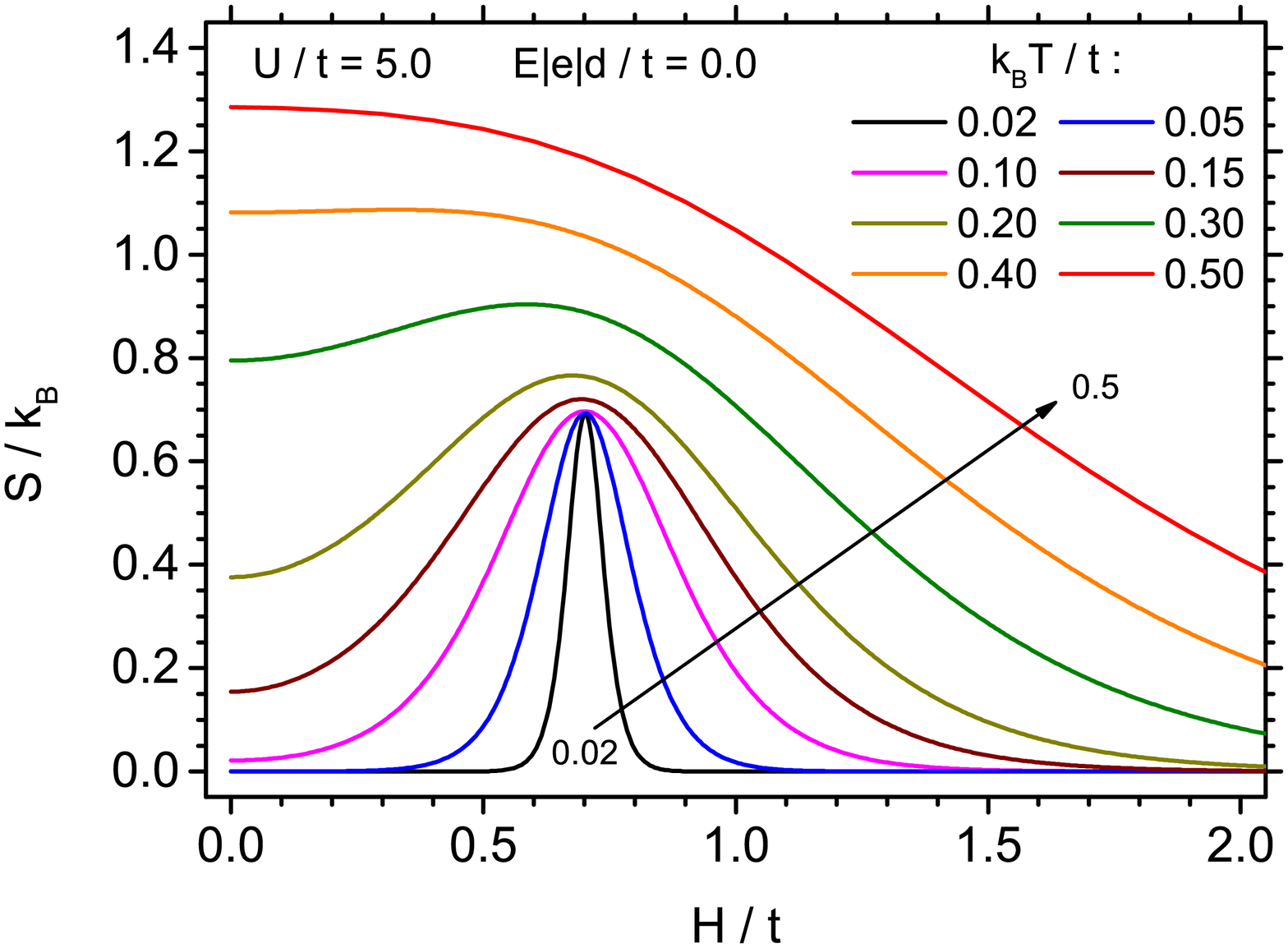}
\caption{\label{fig6}Dependence of the normalized entropy on the normalized magnetic field, for $U/t=5.0$ and electric field $E|e|d/t=0.0$, for various temperatures.}
\end{center}
\end{figure}

Fig.~\ref{fig4} and Fig.~\ref{fig5} present the density and contour plots of the normalized specific heat, $C_{H,E}/k_{\rm B}$, for the same parameters $E|e|d/t$, $H/t$ and $U/t$ as in Fig.~\ref{fig1} and Fig.~\ref{fig2}, respectively. The isolines, representing the constant values of the specific heat, show a quite complex behaviour. The specific heat for $T \to 0$ tends to zero, in agreement with the 3rd law of thermodynamics, in this way reflecting the flattening of the entropy curves from Fig.~\ref{fig3}. The quantum level crossings at $T=0$ are seen in the points where the isolines are concentrating and even forming a loop structure. As the entropy in the vicinity of the level crossing presents a local maximum as a function of the field, the specific heat $C=T\left(\partial S/\partial T\right)$ shows a double peak (with the peaks located at two inflection points of the field dependence of the entropy). With an increase in temperature, in both figures the specific heat increases, then reaches some maximum at intermediate temperatures and finally tends to zero when the temperature is very high. The areas where the specific heat is large, i.e., $C_{H,E}/k_{\rm B}>1$, have been distinguished by various shades of yellow and red colours. It can be deduced from Fig.~\ref{fig4} that the highest maximum of the specific heat occurs at $H/t \to 0$, near the temperature of $k_{\rm B}T/t \approx 0.4$. On the other hand, in Fig.~\ref{fig5}, the highest maximum will occur at the largest electric field ($E|e|d/t \approx5$ in this figure), and for the temperature about $k_{\rm B}T/t \approx 0.7$. Moreover, it can be seen in Fig.~\ref{fig5} that for some electric fields, for instance near $E|e|d/t \approx 2.5$, the double-maximum structure of the specific heat can be predicted when temperature increases. Such an interesting behaviour of the specific heat results from a complicated interplay between the magnetic and electric energy terms. It should be noticed that the specific heat fulfils the inequality $C_{H,E}\ge 0$ for any point in ($H,E,T$)-space, which evidences that the system remains in a stable thermal equilibrium.

\begin{figure}[t]
\begin{center}
\includegraphics[width=0.99\columnwidth]{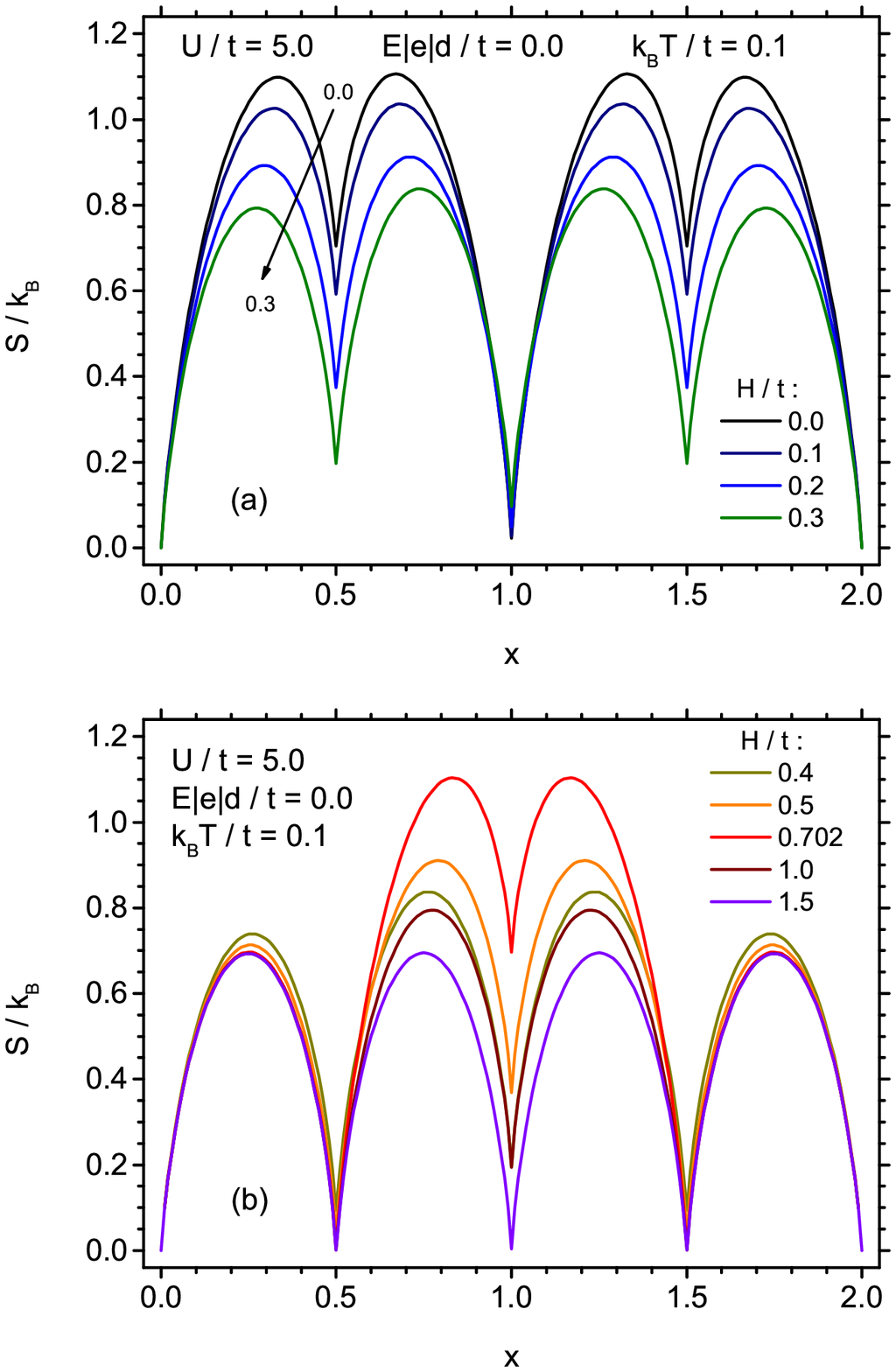}
\caption{\label{fignew2}Dependence of the normalized entropy on the electron concentration, for $U/t=5.0$ electric field $E|e|d/t=0.0$ and normalized temperature $k_{\rm B}T/t=0.1$, for various magnetic fields.}
\end{center}
\end{figure}

\begin{figure}[t]
\begin{center}
\includegraphics[width=0.99\columnwidth]{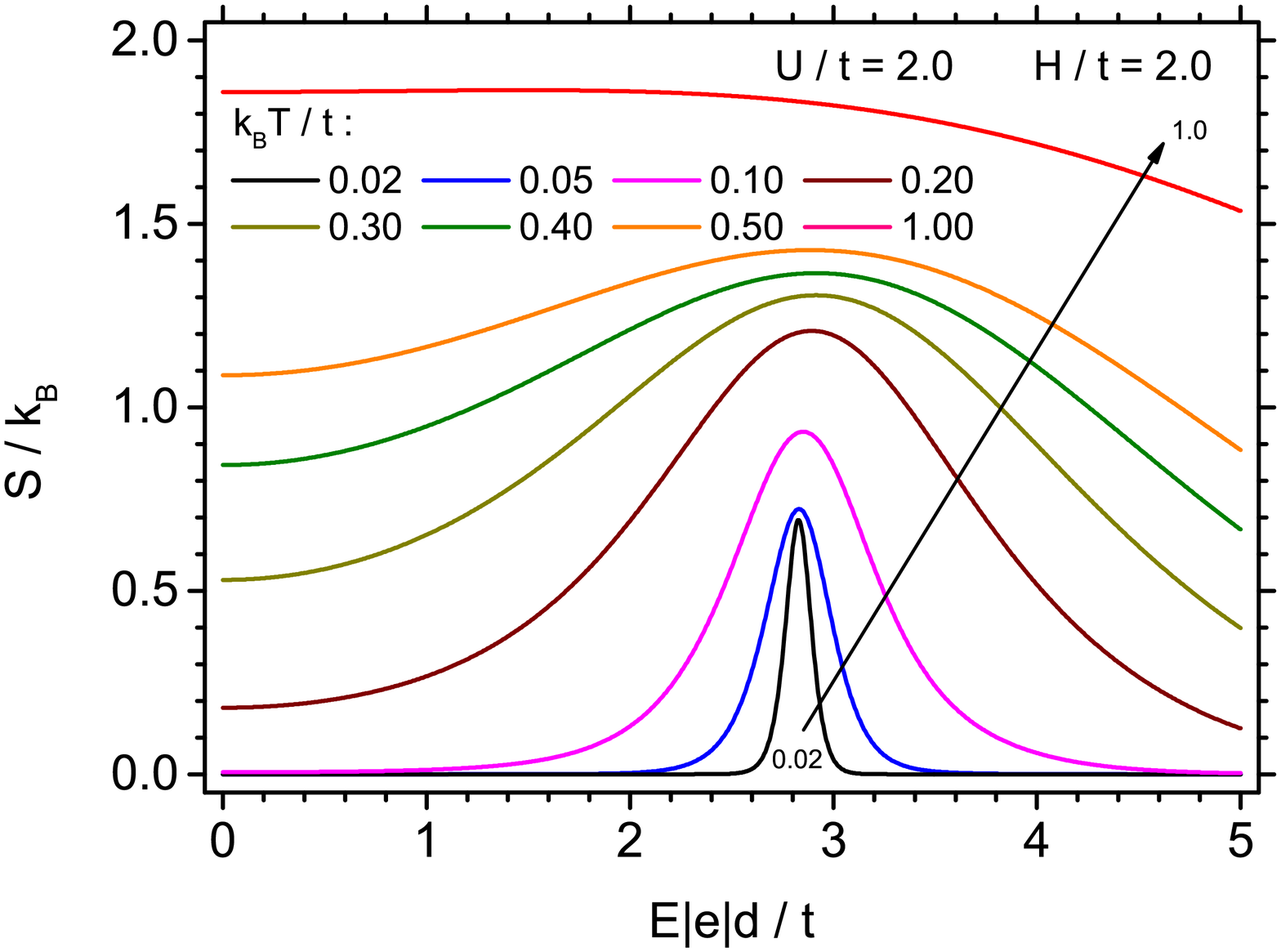}
\caption{\label{fig7}Dependence of the normalized entropy on the normalized electric field, for $U/t=2.0$ and magnetic field $H/t=2.0$, for various temperatures.}
\end{center}
\end{figure}

\begin{figure}[t]
\begin{center}
\includegraphics[width=0.99\columnwidth]{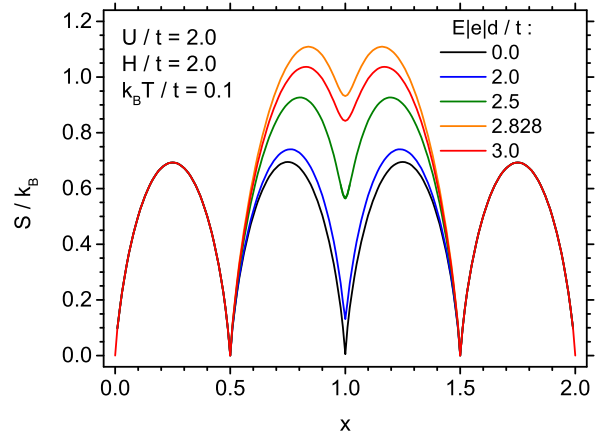}
\caption{\label{fignew1}Dependence of the normalized entropy on the electron concentration, for $U/t=2.0$ magnetic field $H/t=2.0$ and normalized temperature $k_{\rm B}T/t=0.1$, for various electric fields.}
\end{center}
\end{figure}

\begin{figure}[t]
\begin{center}
\includegraphics[width=0.99\columnwidth]{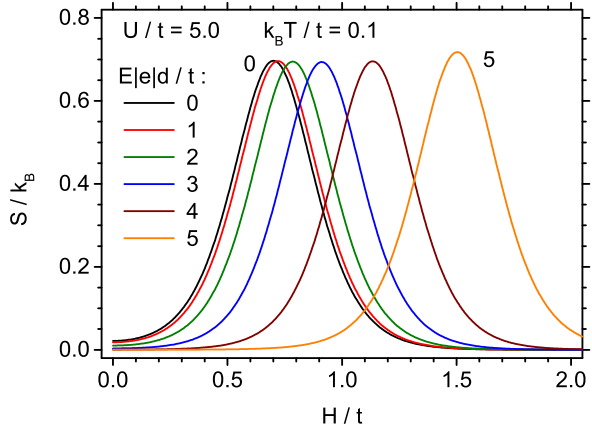}
\caption{\label{fig8}Dependence of the normalized entropy on the normalized magnetic field, for $U/t=5.0$ and normalized temperature $k_{\rm B}T/t=0.1$, for various electric fields.}
\end{center}
\end{figure}

Further illustration of the entropy behaviour is shown in the Fig.~\ref{fig6} to Fig.~\ref{fig8}. In Fig.\ref{fig6} the normalized entropy, $S/k_{\rm B}$, is plotted vs. normalized magnetic field $H/t$. The on-site Coulomb energy amounts to $U/t=5$, whereas the electric field is absent. Various curves correspond to different temperatures. For very low temperatures a peak of residual entropy is seen at the critical magnetic field $H_c$ corresponding to the quantum level crossing of singlet and triplet states. As before, this transition has been predicted by the phase diagram obtained by us in Ref.~\cite{Balcerzak2018e} and the entropy in this doubly-degenerated point for $T \to 0$ amounts to $S/k_{\rm B}=\ln 2$ (see also the discussion of Fig.~\ref{fig3}). The entropy peak disappears in large temperatures, where the curves become monotonously decreasing functions of the field. This result can be intuitively understood, since the increasing magnetic field orders the system and thus diminishes the entropy. Moreover, in the context of Maxwell relation $\left(\partial S/\partial H \right)_{T,E} = \left(\partial M/\partial T \right)_{H,E}$, it means that the magnetization $M$ of the system decreases with the temperature. By the same token, in the region of low temperatures, for the magnetic fields below the quantum level crossing point (in singlet state), the behaviour of magnetization should be anomalous, with the derivative $\left(\partial M/\partial T \right)_{H,E} >0$. Such anomalous behaviour is in agreement with the predictions of our previous paper Ref.~\cite{Balcerzak2018d}.

In order to give a flavour of the importance of electronic concentration on the dimer entropy away from half-filling of the energy levels (i.e. for $x\neq 1$), Fig.~\ref{fignew2} is presented. It permits tracking of the entropy dependence on the electronic concentration $x$ for the same parameters as those used for preparing Fig.~\ref{fig6}, with some values of the external magnetic field selected. In part Fig.~\ref{fignew2}(a) the lower magnetic fields are considered. It is evident that the entropy shows pronounced dependence on the electronic concentration $x$ with full electron-hole symmetry (i.e. the values remain the same for $x$ and $1-x$). In the absence of the magnetic field the entropy is low (close to zero) at half-filling ($x=1$), and if the dimer is charge doped, the entropy rises significantly. The maximum values are taken at $|x-1|=1/3$ and if the system is doped stronger, minimum are reached at $|x-1|=1/2$. Further doping results in reaching another maxima of entropy at $|x-1|=2/3$ and then the entropy tends to zero if the limit of $|x-1|=1$ is achieved (system is empty or completely filled with electrons). A similar behaviour can be observed for the presence of the external field $H/t\lesssim 0.3$, but the entropy value at $x=1$ rises whereas the value at the maxima is reduced. Moreover, the first maximum is shifted towards $x=1$ and the second one shows quite the opposite tendency. If the magnetic field exceeds $H/t\simeq 0.3$, as shown in Fig.~\ref{fignew2}(b), the maxima continue to shift in the directions described above. However, this time the maxima close to $|x-1|\simeq 1/3$ become gradually more pronounced. The largest entropy value at the maxima is achieved at the critical magnetic field $H/t=0.702$. After crossing this field value, the maxima tend to flatten. On the contrary the maxima close to $|x-1|\simeq 2/3$ exhibit the entropy magnitude much less sensitive to the magnetic field for $H/t\gtrsim 0.3$; when it increases, the value tends to $k_{\rm B}\ln 2$ for strong fields and the position of maxima shifts to $|x-1|=3/4$.

Behaviour of the entropy analogous to one discussed in Fig.~\ref{fig6} can be seen in Fig.~\ref{fig7}, where it is plotted vs. the electric field $E|e|d/t$, for Coulomb on-site energy $U/t=2$ and the magnetic field fixed at $H/t=2$. Various curves correspond to different temperatures. Again, in the low temperature region, the residual entropy peak is seen, corresponding to quantum level crossing between the triplet and the singlet state. The value of the entropy in the peak for $T \to 0$ is the same as in Fig.~\ref{fig6}, $S/k_{\rm B}=\ln 2$. The peak disappears as the temperature increases. The analysis of the entropy vs. electric field can be connected with the behaviour of electric polarization $P$ vs. temperature. With the help of Maxwell relation $\left(\partial S/\partial E \right)_{H,T} = \left(\partial P/\partial T \right)_{H,E}$, we can conclude that the electric polarization $P$ should behave similarly to magnetization $M$. Namely, it decreases with increase in temperature for sufficiently large temperatures. However, in the low temperatures, below the quantum level crossing (in triplet state) the behaviour of $P$ is anomalous, with the derivative $\left(\partial P/\partial T \right)_{H,E} >0$. Such a behaviour has also been predicted in Ref.~\cite{Balcerzak2018d}.

Like in the case of Fig.~\ref{fignew2}, it is instructive to sketch the entropy behaviour as a function of the electron concentration to demonstrate the effect of charge doping of the system. Such data can be tracked in Fig.~\ref{fignew1}, prepared for the same parameters as Fig.~\ref{fig7}, for selected values of the electric field. In the absence of the electric and magnetic field, the entropy shows four symmetric maxima at $|x-1|=1/3$ and $|x-1|=2/3$, with the entropy value of $k_{\rm B}\ln 2$. At $|x-1|=1/2$ the entropy has deep local minima. The maxima at $|x-1|=2/3$ remain completely insensitive to the changes in the electric field. On the contrary, the maxima at $|x-1|=1/3$ build up when the electric field is applied; their position is shifted towards $x=1$. Also the minimum at $x=1$ is lifted up. The maximum entropy is reached at the critical electric field value $E|e|d/t=2.828$. Further increase in the field results in flattening of the maxima. 

Both Fig.~\ref{fignew2} and Fig.~\ref{fignew1} demonstrate the crucial influence of the electronic concentration on the entropy behaviour in the dimer. However, the case of $x=1$ (i.e. half filling), indicates the most pronounced sensitivity of the entropy to the external electric field. Therefore, we return in all the further discussion to the case of $x=1$.

The observation that maximum value of the entropy in the low-temperature peak is constant, $S/k_{\rm B}=\ln 2$, has been confirmed in Fig.\ref{fig8}. It this figure the normalized entropy, showing a pronounced peak, is presented vs. normalized magnetic field $H/t$, for $U/t=5$, whereas the temperature is low and constant, $k_{\rm B}T/t =0.1$. Various curves in Fig.\ref{fig8} correspond to different magnitudes of the electric field $E|e|d/t$. An increase in the electric field causes the increase in the critical magnetic field $H_c$, thus shifting the position of the entropy peak. Such a shift is non-linear vs. electric field $E|e|d/t$, which is in agreement with the phase diagram presented in Ref.~\cite{Balcerzak2018e}.

\begin{figure}[t]
\begin{center}
\includegraphics[width=0.99\columnwidth]{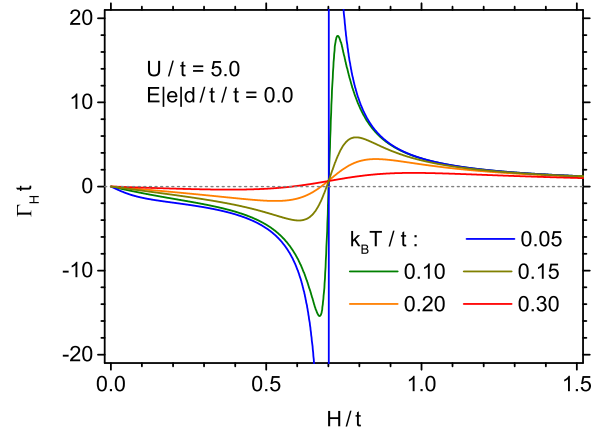}
\caption{\label{fig9}Dependence of the normalized magnetic Gr\"uneisen ratio on the normalized magnetic field, for $U/t=5.0$ and normalized electric field $E|e|d/t=0.0$, for various temperatures.}
\end{center}
\end{figure}

The response of the entropy to the external fields is quantified by the appropriate Gr\"uneisen ratios. In Fig.~\ref{fig9} the magnetic Gr\"{u}neisen ratio $\Gamma_H t$ is shown in dimensionless units, as a function of the normalized magnetic field $H/t$. Various curves correspond to different temperatures $k_{\rm B}T/t$. The Coulombic on-site energy amounts to $U/t=5$, and the electric field is absent. It is demonstrated that for very low temperatures the parameter $\Gamma_H$ diverges at the quantum level crossing, i.e., for the critical field $H_c$. With an increase in temperature, the divergence disappears and the curves flatten. The magnetic Gr\"{u}neisen ratio is negative in the range of the singlet ground state, i.e., below $H_c$, and positive in the range of the triplet ground state, above $H_c$. The divergence of $\Gamma_H$ at $H_c$, when $T \to 0$, is predisposing this parameter for a good indicator of the quantum level crossing.

\begin{figure}[t]
\begin{center}
\includegraphics[width=0.99\columnwidth]{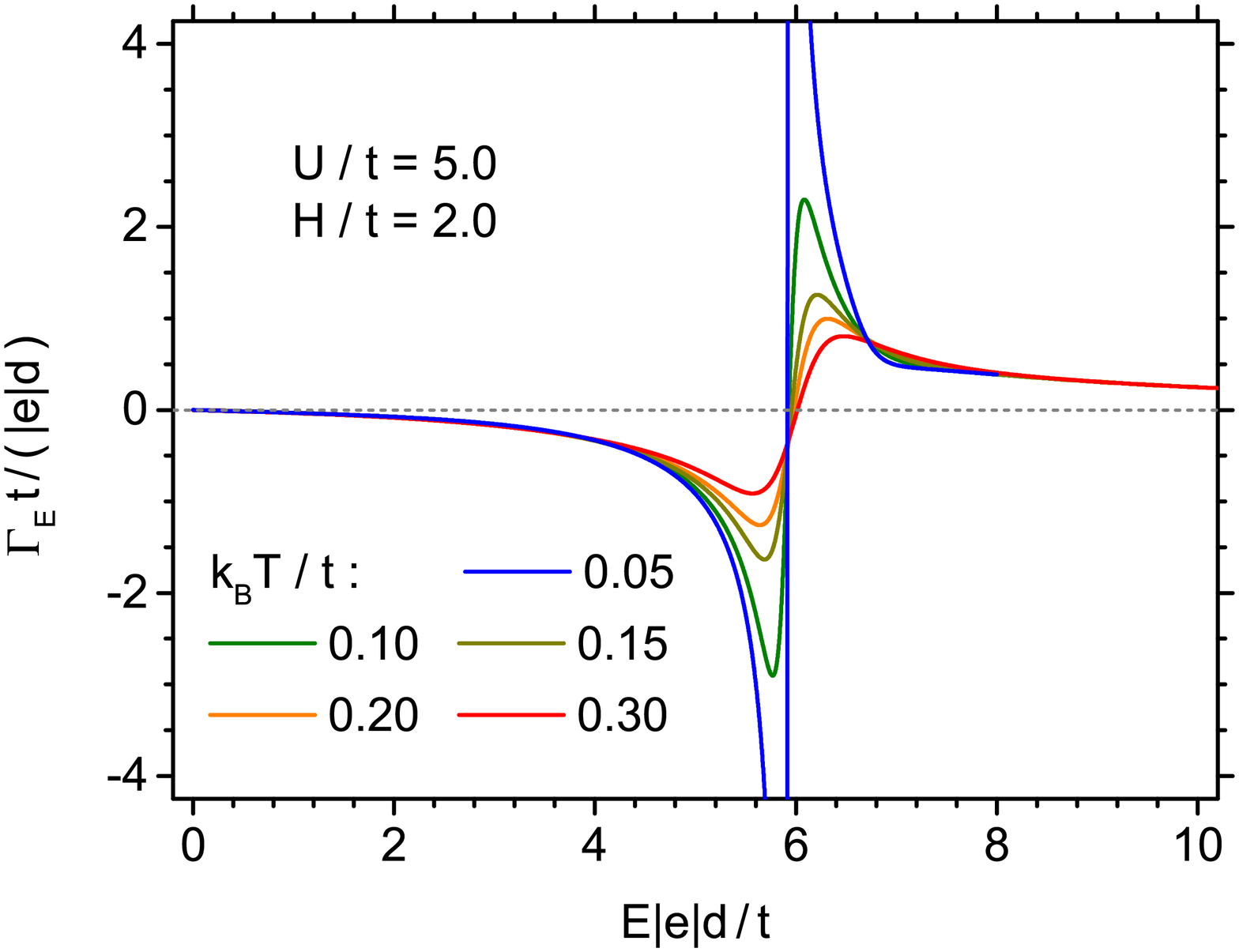}
\caption{\label{fig10}Dependence of the normalized electric Gr\"uneisen ratio on the normalized electric field, for $U/t=5.0$ and normalized magnetic field $H/t=2.0$, for various temperatures.}
\end{center}
\end{figure}

In Fig.~\ref{fig10} the electric Gr\"{u}neisen ratio, $\Gamma_E t/\left(|e|d\right)$, is shown in dimensionless units, vs. normalized electric field $E|e|d/t$. Various curves correspond to different temperatures $k_{\rm B}T/t$. The Coulombic on-site energy amounts to $U/t=5$, and the magnetic field is $H/t=2$. One can see in Fig.~\ref{fig10} that for very low temperatures the parameter $\Gamma_E$ diverges at the quantum level crossing point, at the critical field $E_c$. With an increase in temperature this divergence disappears and the curves flatten, similarly to the behaviour of $\Gamma_H$ demonstrated in the previous figure. However, here the electric Gr\"{u}neisen ratio is negative in the range of the triplet ground state, i.e., below $E_c$, and is positive in the range of the singlet ground state, above $E_c$. The divergence of $\Gamma_E$ at $E_c$, when $T \to 0$, can be also a useful property, analogously to $\Gamma_H$, for uncovering the presence of the quantum level crossing. Moreover, this divergence proves the sensitivity of the entropy to the changes of the external field, marking the regions of interest for maximising the entropy change in the caloric effects.

\begin{figure}[t]
\begin{center}
\includegraphics[width=0.99\columnwidth]{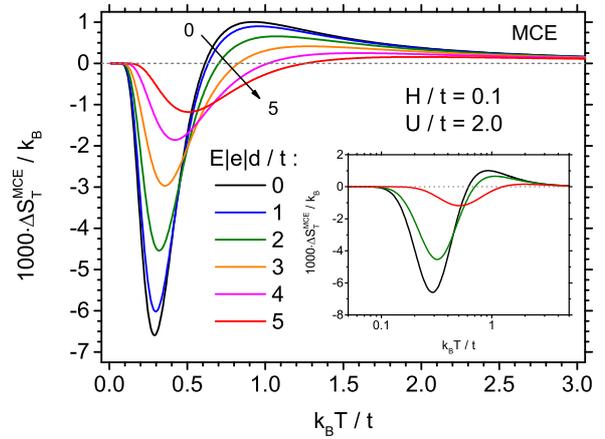}
\caption{\label{fig11}Dependence of the normalized isothermal entropy change in magnetocaloric effect on the normalized temperature, for magnetic field variation between $H/t=0.0$ and 0.1, for $U/t=2.0$ and various normalized electric field values. The inset shows selected data from the main panel in logarithmic scale for temperature.}
\end{center}
\end{figure}

\begin{figure}[t]
\begin{center}
\includegraphics[width=0.99\columnwidth]{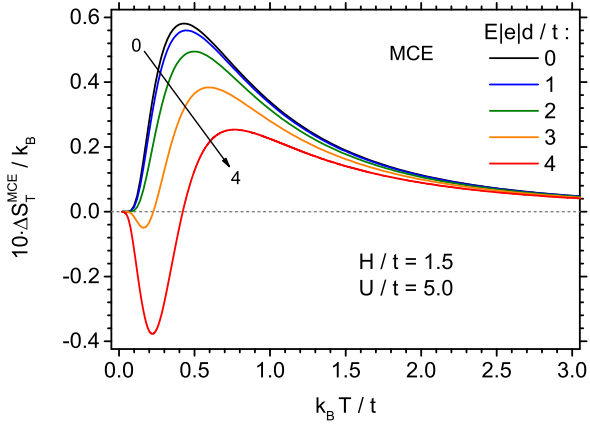}
\caption{\label{fig12}Dependence of the normalized isothermal entropy change in magnetocaloric effect on the normalized temperature, for magnetic field variation between $H/t=0.0$ and 1.5, for $U/t=5.0$ and various normalized electric field values.}
\end{center}
\end{figure}

The behaviour of the entropy as a function of the external field is crucial for the description of magnetocaloric (MCE) and electrocaloric (ECE) effects. The appropriate quantity is the isothermal entropy change when the external field is varied between the initial value of 0 to the non-zero final value. First let us discuss the MCE in the Hubbard dimer system. In Fig.~\ref{fig11} and Fig.~\ref{fig12} the isothermal entropy change in MCE (defined by Eq.~\ref{deltaSH}) is presented vs. normalized temperature $k_{\rm B}T/t$. Various curves in these figures correspond to different constant external electric fields $E|e|d/t$. Fig.~\ref{fig11} is prepared for $U/t=2$ and the external magnetic field is switched from $H/t=0$ to $H/t=0.1$. The final value, according to the phase diagram (Ref.~\cite{Balcerzak2018e}), corresponds to the region of the singlet ground state for all the values of the electric field. On the other hand, Fig.~\ref{fig12} is prepared for $U/t=5$ with the magnetic field switched from $H/t=0$ to $H/t=1.5$, where such a final value corresponds to the triplet ground state for the values of electric fields specified in the figure legend. It can also be mentioned that for the electric field higher than about $E|e|d/t \approx 5$ the system with the parameters from Fig.~\ref{fig12} should undergo the transition to the singlet ground state. One can see that behaviour of the MCE curves vs. temperature is completely different for both figures. In Fig.~\ref{fig11}, starting from the singlet ground state, the strong inverse MCE can be observed in the low temperature region. Then, for higher temperatures the MCE becomes direct, i.e., entropy change is positive, but relatively weak, and it further weakens with an increase in temperature. The inset in Fig.~\ref{fig11} shows this behaviour in the logarithmic temperature scale. On the other hand, in Fig.~\ref{fig12}, starting from the triplet ground state, a strong direct MCE can be observed, provided the electric field is not very large. When $E|e|d/t$ increases, that is, approaching a phase boundary with the singlet ground state, the inverse MCE appears in the low temperatures and the curves noticeably tend to the shape demonstrated in the previous figure. The common characteristic of both figures, Fig.~\ref{fig11} and Fig.~\ref{fig12}, is a shift of the minimum and maximum position of the curves towards higher temperatures, as the electric field increases. The strengthening of the electric field makes the curves more flat in Fig.~\ref{fig11}, but in Fig.~\ref{fig12} such conclusion can be drawn only for the positive (direct) MCE. It can be seen that the magnitude of the external constant electric field exerts a noticeable effect on the entropy change under variation of the magnetic field, being a clear manifestation of the magnetoelectric phenomena in Hubbard dimer.

\begin{figure}[t]
\begin{center}
\includegraphics[width=0.99\columnwidth]{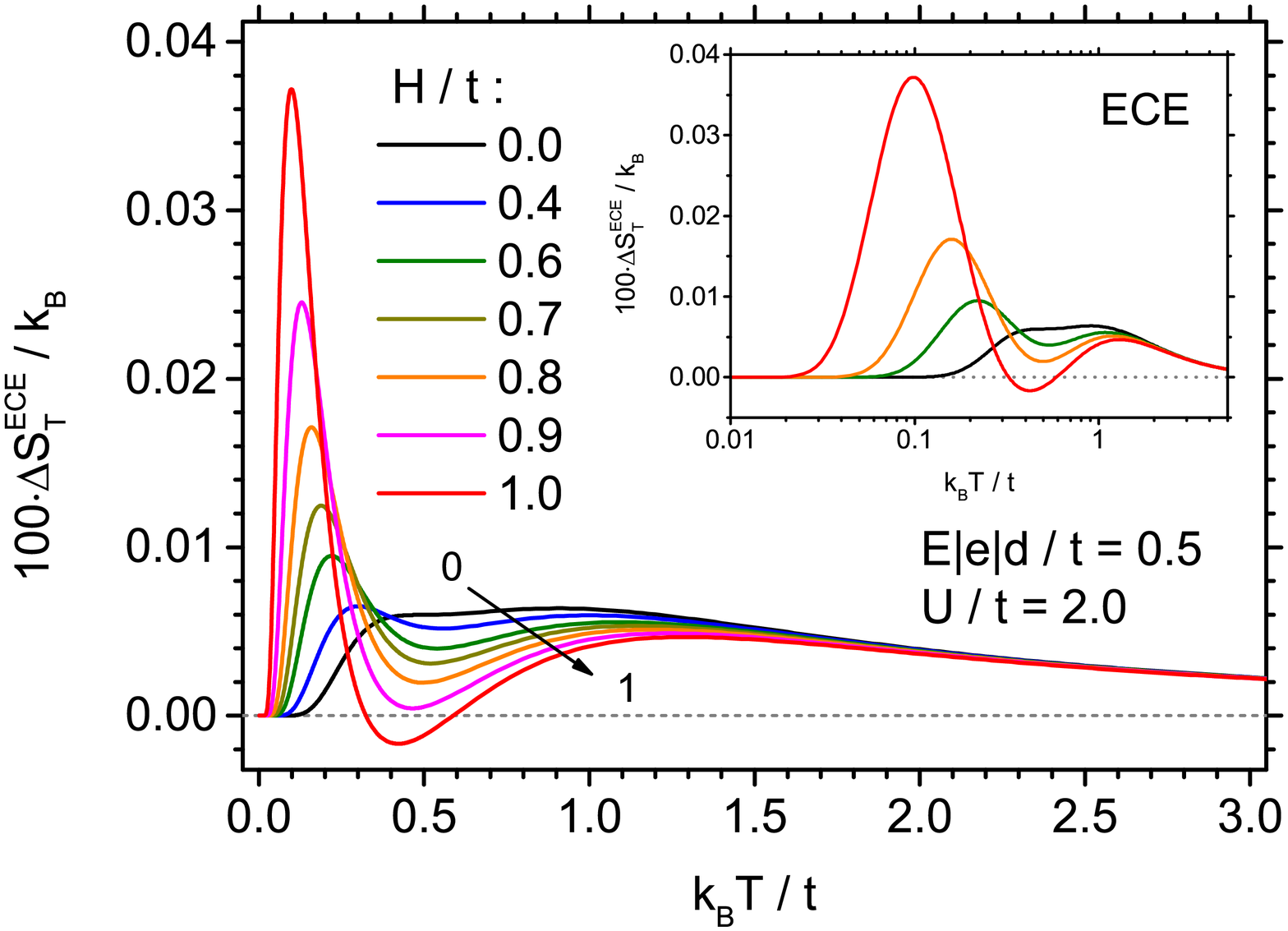}
\caption{\label{fig13}Dependence of the normalized isothermal entropy change in electrocaloric effect on the normalized temperature, for electric field variation between $E|e|d/t=0.0$ and 0.5, for $U/t=2.0$ and various normalized magnetic field values. The inset shows selected data from the main panel in logarithmic scale for temperature.}
\end{center}
\end{figure}

\begin{figure}[t]
\begin{center}
\includegraphics[width=0.99\columnwidth]{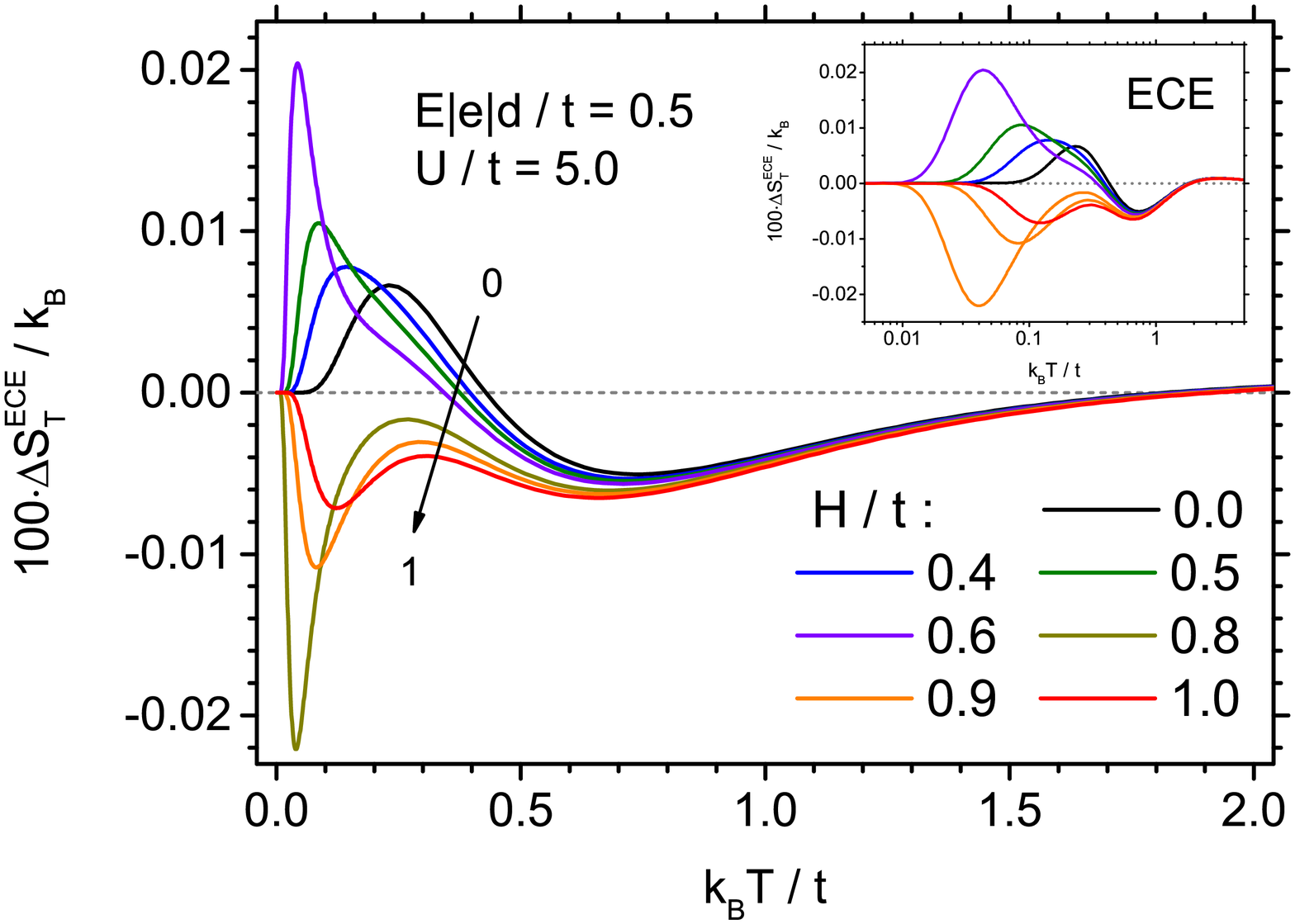}
\caption{\label{fig14}Dependence of the normalized isothermal entropy change in electrocaloric effect on the normalized temperature, for electric field variation between $E|e|d/t=0.0$ and 0.5, for $U/t=5.0$ and various normalized magnetic field values. The inset shows selected data from the main panel in logarithmic scale for temperature.}
\end{center}
\end{figure}

An effect complementary to MCE is ECE, quantified conveniently by the isothermal entropy change defined by Eq.~\ref{deltaSE}. In Fig.~\ref{fig13} and Fig.~\ref{fig14}, the isothermal entropy change in ECE is presented vs. normalized temperature $k_{\rm B}T/t$. Various curves in these figures correspond to different magnitudes of the constant external magnetic field $H/t$. Fig.~\ref{fig13} is prepared for $U/t=2$ and the electric field is switched from $E|e|d/t=0$ to $E|e|d/t=0.5$, where the final value corresponds to the region with the singlet ground state, since all values of the magnetic field labelled in this figure fulfil the condition $H<H_c$. On the other hand, Fig.~\ref{fig14} is prepared for $U/t=5$ and the electric field is also switched from $E|e\d/t=0$ to $E|e|d/t=0.5$. Therefore, the final value of $E$ corresponds to the singlet ground state for $H/t=$0.0, 0.4, 0.5 and 0.6, according to the phase diagram presented in Ref.~\cite{Balcerzak2018e}. However, for higher fields, i.e., for $H/t=$0.8, 0.9 and 1.0, the triplet ground state occurs at the final value of $E$. Therefore, it is interesting to compare ECE in both these figures.
In Fig.~\ref{fig13} the curves present two positive maxima showing a direct ECE. In the low temperature region the maximum is sharp and it builds up with an increase in the magnetic field. On the other hand, the second, high-temperature maximum is less pronounced and slowly flattens with an increase in the magnetic field. At the same time, the minimum between these maxima becomes deeper as the field increases. In particular, for $H/t=1$ an inverse ECE can be found, corresponding to the minimum of $\Delta S_T^{ECE}$, which then extends down to negative values. In Fig.~\ref{fig14}, the ECE curves corresponding to $H/t\le 0.6$, i.e., for the singlet ground state, are of similar character as those in Fig.~\ref{fig13}, with the reservation that the negative minima are much deeper. However, the curves for $H/t\ge 0.8$, i.e., for the triplet ground state, are totally different. They present a strong inverse ECE in the low-temperature minimum. Also the second negative minimum, which is more shallow, can be seen for larger temperatures. The minimum at the low temperature is especially pronounced for $H/t=0.8$, i.e., for the smallest considered field in the range of the triplet ground state. Evidently, the rapid change from positive to negative ECE is connected with the quantum phase transition in the ground state, from singlet to triplet state.  
The insets in Figs.\ref{fig13} and \ref{fig14} are to inspect the effect in the logarithmic temperature scale for some representative values of $H/t$ chosen from the main figures, to facilitate tracking the low-temperature behaviour.

%section 4
\section{\label{final}Summary and conclusion}

In the paper we report a theoretical study of the caloric effects - MCE and ECE - in a model Hubbard dimer (pair cluster) immersed in external electric and magnetic field.

The formalism of the grand canonical ensemble has been used \cite{Balcerzak2017c}, in which the system can exchange electrons with its environment, whereas the average electron concentration, $x=\left(\left<n_a\right>+\left<n_b\right>\right)/2$, amounts to $0\le x \le 2$. This general formalism enables the studies of the influence of electron concentration $x$ on the thermodynamic properties of such cluster.
However, in the numerical application of the method, the most interesting value of concentration has been exploited, namely $x=1$, which corresponds to the half-filling of the energy levels of the system. For $x=1$ the sensitivity of the system entropy to the external fields was found to be maximized, thus corresponding to the most pronounced caloric effects exhibited by the system in question.
 
For investigation of the caloric effects, a crucial quantity is the entropy with its dependence on the external fields. The numerically exact results for the entropy of the Hubbard dimer have been analysed in relation to the phase diagram obtained by us previously in Ref.~\cite{Balcerzak2018e}, and they are in agreement with other thermodynamic properties, for instance, those calculated in our work Ref.~\cite {Balcerzak2018d}. One of the most interesting results is the residual entropy in the ground state exactly at the quantum level crossing point (corresponding to a transition between the singlet and triplet state and controlled with the electric or magnetic field). This particular feature is manifested as the finite-width entropy peak at the finite temperature when the critical field value is crossed. Moreover, in the vicinity of this critical field the entropy is particularly sensitive to the external field, thus maximizing the caloric effect magnitude. It might be mentioned that the effect of maximization of the field dependence of the entropy has been pointed out in the context of spin dimers \cite{Sharples2014,Chakraborty2019} and the singlet-triplet transition in this system focused the attention in Ref.~\cite{Chakraborty2015}.

The numerical calculations concerned first measures of MCE and ECE such as the isothermal entropy changes $\Delta S_T^{MCE}$ and $\Delta S_T^{ECE}$, respectively. The investigations spanned of the wide range of temperatures as well as magnitudes of the magnetic field and electric field change, to identify the most interesting cases. In general, the significant ranges of both direct and inverse caloric effects were found. 

It has also been found that the caloric effects are especially pronounced in the low temperature region, in the vicinity of the critical fields responsible for quantum level crossing. This fact could be potentially used in practice, for the magnetic or electric field change-based cooling in the range of low temperatures.

In the theoretical part, the electric Gr\"{u}neisen ratio, $\Gamma_E$, has been defined as a new quantity, being an analogue of the magnetic Gr\"{u}neisen ratio, $\Gamma_H$. It has been shown that both parameters reveal singularities at the quantum critical points, when $T \to 0$. Thus, both these Gr\"{u}neisen ratios can be useful for determination of the quantum phase transitions.

The Hubbard dimer (pair cluster) turned out to be a very interesting model system, in which the thermodynamics of the caloric effects - MCE and ECE - can be simultaneously studied by the exact method. The magnetic and electric fields are found to exert opposite effects on the induced magnetism in the studied system, therefore, using the exact approach is particularly important for studying the interplay of both fields.

It can be noted that in the light of the paper \cite{Balcerzak2019a}, where the magnetoelastic properties of the Hubbard pair cluster have been investigated, it would be interesting to study the caloric effects in the presence of the external elastic forces , leading to the emergence of multicaloric effects \cite{Ursic2016,Stern-Taulats2018,Hao2020}. These forces, being able to deform interatomic distance, influence both the hopping integral and the interatomic Coulomb potential. However, such problem exceeds the frame of the present paper and should be considered elsewhere.

Having found the entropy, the specific heat has been determined for the system embedded simultaneously in the magnetic and electric fields. An interesting result is a possibility of occurrence of double maximum of the specific heat in some regions of the density diagrams (Figs.~\ref{fig4} and \ref{fig5}), when corresponding temperature dependence is analysed.

The dimer with electron hopping has also been discussed in the literature as a part of Ising Hamiltonian-based more elaborate magnetic model \cite{Cencarikova2018a,Cencarikova2019,Cencarikova2020a} in the simultaneous presence of the electric and magnetic field. In this case, non-trivial phase diagrams were found as a result of the interplay of the charge doping (controlled with the chemical potential) and the influence of the external fields in the infinite planar system. This might motivate further studies of Hubbard dimers embedded in localized-spin magnetic model in the external fields.

Last but not least, let us mention the applicability of the Hubbard dimer model to experimental systems in condensed matter physics. In this context, we can point out the layered charge transfer salts of Mott insulator type, for which Hubbard dimer can serve as a minimum model \cite{McKenzie1999}. For example, the extended Hubbard model for two sites was investigated to capture physics of $\beta$ and $\kappa$ polymorphs of (ET)$_{2}$X or (BEDT-TTF)$_{2}$X, where (ET) or (BEDT-TTF) is bis(ethylenedithio)tetrathiafulvalene and X is monovalent anion, like Cu$_2$(CN)$_3$, constituting half-filled systems composed of effective dimers \cite{Scriven2009}. Similar goals was addressed in calculations performed in Ref.~\cite{Koretsune2014}. The mentioned materials motivate also the interest in development of more elaborate models based on dimerized Hubbard Hamiltonian with interacting dimers \cite{Jacko2020} for a complex group of BEDT-TTF charge transfer salts \cite{Dressel2020}. A somehow similar kind of model has been applied to description of metal-insulator transition in VO$_2$ \cite{Najera2017}. In this context, it is worth mentioning that the pronounced electrocaloric effect associated with metal-insulator transition has been measured in VO$_2$ \cite{Matsunami2015}. In addition, etracyanoquinodimethane (TCNQ)-based charge transfer salts were also modelled using the dimer Hubbard model \cite{Kral1980}. Also, the applications of two site Hubbard model to description of diradicals can be mentioned \cite{Calzado2002,Nakano2007,Kamada2010}. Hubbard model on small clusters has been also invoked for prediction of selected properties of transition metal nanostructures \cite{Lopez-Urias2005,Lopez-Urias2009}.

The present results, concerning an ensemble of non-interacting Hubbard dimers, may serve as a starting point for the studies of electro- and magnetocaloric phenomena in extended model with, for example, interdimer interactions included (see for example the study in Ref.~\cite{Fye1992}). Moreover, larger clusters can be studied using the identical model, to mention for example our work on the cubic cluster Ref.~\cite{Szalowski2018c}. The influence of the cluster geometry on the thermodynamic properties is expected to be crucial (in particular, linear or closed geometry can lead to different sort of behaviour, as it can be followed for the case of Hubbard trimer in Ref.~\cite{Juliano2019}). Therefore, each shape and size of cluster requires a separate computation. In the context of exact studies, the works for various tetramers can be noticed \cite{Schumann2002,Schumann2007a,Schumann2008a}. However, in relation to the external electric field applied to the Hubbard model for large clusters or even infinite lattice of any dimensionality, it should be mentioned that screening effects would limit the field influence on the model properties. This facts focuses the interest in electric field-related phenomena rather on small clusters.

%\section{References}

%\bibliographystyle{elsarticle-num}
%\bibliography{cool}

\end{document}